\documentclass[12pt]{article}
\usepackage{graphicx}
\usepackage{bm}        % for math
\usepackage{amssymb}   % for math
\usepackage{hyphenat}

\usepackage{ textcomp }
\input epsf
\def\be{\begin{eqnarray}}
\def\ee{\end{eqnarray}}
\def\beq{\begin{eqnarray}}
\def\eeq{\end{eqnarray}}

\def\K{\mathcal{K}}

\def\ba{\begin{eqnarray}}
\def\ea{\end{eqnarray}}
\def\beq{\begin{eqnarray}}
\def\eeq{\end{eqnarray}}

\def\mpl{M_{\rm Pl}}

\def\d{\mathrm{d}}
\def\p{{\cal P}}

\def\L*{{\cal L}_*}
\def\L{\mathcal{L}}
\def\({\left(}
\def\){\right)}
\def\ie{{\it i.e. }}
\def\eg{{\it e.g. }}
\def\nn{\nonumber}
\def\p{\partial}
\def\mn{_{\mu \nu}}
\def\stu{St\"uckelberg }
\def\p{\partial}

\def\<{\langle}
\def\>{\rangle}

\def\be{\beta}
\def\lsim{\mathrel{\rlap{\lower3pt\hbox{\hskip0pt$\sim$}}
     \raise1pt\hbox{$<$}}}         %less than or approx. symbol
\def\gsim{\mathrel{\rlap{\lower4pt\hbox{\hskip1pt$\sim$}}
     \raise1pt\hbox{$>$}}}         %greater than or approx. symbol

\def\lsim{\mathrel{\rlap{\lower3pt\hbox{\hskip0pt$\sim$}}
     \raise1pt\hbox{$<$}}}         %less than or approx. symbol
\def\gsim{\mathrel{\rlap{\lower4pt\hbox{\hskip1pt$\sim$}}
     \raise1pt\hbox{$>$}}}         %greater than or approx. symbol
\def\stu{St\"uckelberg }

\usepackage{amsmath}
\usepackage{amsfonts}
\usepackage{verbatim}
\usepackage{setspace}
\usepackage{url}
\usepackage{color}
\usepackage{mcite}
\begin{document}

%\maketitle

\begin{flushright}
{UCSD-PTH-12-20}

{NYU-TH-11/11/12}
\end{flushright}

\begin{center}
{\LARGE \bf{Non-Renormalization and Naturalness }}
\vskip 0.5cm
{\LARGE \bf{in a Class of Scalar-Tensor Theories}}
\end{center}
%\vspace{1truecm}
\thispagestyle{empty}
\begin{center}
{\large  { Claudia de Rham${}^{a}$,}}
{\large  { Gregory Gabadadze${}^{b}$,}}
{\large  { Lavinia Heisenberg${}^{a,c}$ and}}
{\large  {David Pirtskhalava${}^{d}$}}

 \vspace{0.5 cm}

{\it ${}^a$Department of Physics, Case Western Reserve University,
10900 Euclid Ave, Cleveland, OH
44106, USA}

\vspace{.2cm}
{\it $^b$Center for Cosmology and Particle Physics,
Department of Physics,}
\centerline{\it New York University, New York,
NY, 10003, USA}

\vskip 0.2cm

{\it $^c$D\'epartement de Physique Th\'eorique and Center for Astroparticle Physics,
 Universit\'e de Gen\`eve, 24 Quai E. Ansermet, CH-1211,  Gen\`eve, Switzerland}

 \vspace{.2cm}

{\it $^d$Department of Physics, University of California, San Diego,\\ La Jolla,  CA 92093 USA}

\end{center}

\begin{abstract}

We study the renormalization of some dimension-4, 7 and 10 operators  in a class of nonlinear scalar-tensor theories. These theories are invariant under: (a) linear diffeomorphisms which represent an exact symmetry of the full non-linear action, and (b) global field-space Galilean transformations of the scalar field. The Lagrangian contains a set of
non-topological interaction terms of the above-mentioned dimensionality, which we show are not renormalized at  any order in perturbation theory. We also discuss the renormalization  of other operators,
that  may be generated by loops  and/or receive loop-corrections, and identify  the
regime in which they are sub-leading with respect to the operators that do not get renormalized.
Interestingly, such scalar-tensor theories emerge  in a certain high-energy limit of the ghost-free theory of massive gravity. One can use the non-renormalization properties of the high-energy limit to estimate the magnitude of quantum corrections in the full theory. We show that the quantum corrections to the three free parameters of the model, one of them being the graviton mass, are strongly suppressed. In particular, we show that having  an arbitrarily small graviton mass is
technically natural.
\end{abstract}

\newpage

\section{Motivation}

Arguably, two of the biggest puzzles of modern cosmology remain the origin of the
accelerated expansion of the present-day Universe,  and the old cosmological
constant (CC) problem, arising from a giant mismatch between the theoretically expected magnitude
of the vacuum energy, and the tiny value of the observed space-time
curvature. Although the resolution of these two puzzles may be related to each other,
General Relativity (GR) fails to address the CC problem,
while only being able to accommodate cosmic acceleration via the
postulated dark energy, offering no insights into its origin.

Theories that extend/modify GR at a large distance scale (say at some scale $m^{-1}\sim
H_0^{-1}$, $H_0$ denoting the present-day value of  the Hubble parameter)
offer a hope to cancel the vacuum energy by evading S. Weinberg's
no-go theorem, at the same time describing the accelerated expansion
in terms of a small dimensionful parameter $m$.  Although
a satisfactory framework does not yet exist, the resolution
seems to be not too far in the future \cite{deRham:2010tw}.

In such a scenario, a natural question arises as to whether the introduced
small parameter itself (e.g., the graviton mass $m$) is subject to  strong
renormalization by quantum loops, similar to the renormalization  of
a small cosmological constant\footnote{To be clear, by strong
renormalization of a parameter we mean an additive renormalization
which is proportional to positive powers of the UV cutoff, as opposed to
a multiplicative renormalization that is only logarithmically sensitive to it; see
more on this below.}. This is one of the questions
we will address in the present work.

Technically natural tunings are not uncommon within the Standard Model of particle physics. According to 't Hooft's naturalness argument \cite{'tHooft:1979bh,Dimopoulos:1979es}, a physical parameter $c_i$ can remain naturally small at any energy scale $E$ if the limit $c_i\to0$ enhances symmetry of the system.  While the electron mass $m_e$ is much smaller than the electroweak scale for instance, it is technically natural, since quantum corrections give rise to a renormalization of $m_e$ proportional to $m_e$ itself, making the mass parameter logarithmically sensitive to
the UV scale. The reason for this is simple: taking the $m_e\to0$ limit implies an additional chiral symmetry, enforcing the separate conservation of left- and right-handed electrons, meaning that in the massless limit the electron mass receives no quantum corrections.
For the case of the cosmological constant,  $\lambda$, on the other hand, there is no
symmetry recovered in the limit $\lambda\to 0$,  and any particle of mass $M$ is expected to
contribute to the CC by  terms among which are
$M^2 \Lambda_{UV}^2/\mpl^2$, $\Lambda_{UV}^4/\mpl^2$, with $\Lambda_{UV}$ denoting
the UV cutoff. The smallness of the CC is thus
unnatural in the 't Hooft sense.

We will show in the present work, that the introduction of the small graviton mass
$m$ is technically  natural, in sharp distinction from the small CC  scenario.

%The new small scale of an infrared (IR) modification of gravity
%(e.g., the graviton mass $m$), has proven to be
%useful in reproducing the cosmological properties of the observed
%Universe \cite{Deffayet:2000uy,Deffayet:2001pu}.
%Having the small  parameter, $m$,  describing the accelerated
%Universe in a technically natural way, gives this scenario
%a theoretical advantage over the one in which dark energy
%is due to a technically-unnatural small CC.  As an additional huge bonus,
%IR modified gravity in general, and massive gravity
%in particular,  turns out to offer ways to begin to address the old CC
%problem \cite{Dvali:2002pe,Dvali:2002fz,ArkaniHamed:2002fu,Dvali:2007kt}.

A Lorentz-invariant modification of GR, such as massive gravity, introduces extra propagating degrees of
freedom, as well as a  strong coupling energy scale $\Lambda_3$ associated with these degrees of freedom. Usually, the energy scale $\Lambda_3$
is much smaller than the Planck mass, while significantly exceeding $m$. For example, on flat space $\Lambda_3=(\mpl m^2)^{1/3}$; however it can be much higher on non-trivial backgrounds. On the one hand, the presence of strong coupling is typically required to hide unneeded extra forces from observations
via the Vainshtein mechanism \cite{Vainshtein:1972sx,Deffayet:2001uk}
\footnote{Note however,
that a different mechanism of hiding the extra degrees of freedom has been found in \cite {deRham:2010tw},
which does not rely on the Vainshtein effect, but is perturbative. Its
virtues will be discussed later.}. On the other hand, the strongly coupled behavior
calls for important questions on calculability, quantum consistency, and in
this particular case on superluminality, of the massive theory. The first two
are the questions that we will be addressing below. The question of superluminal propagation is tied to that of potential UV extensions of these theories above the scale $\Lambda_3$ \cite{Adams:2006sv,Burrage:2011cr}, and will be addressed elsewhere.

\section{The Effective Field Theory}

We will consider a theory of a  massless spin-two field $h_{\mu\nu}$,
and a scalar $\pi$, which couple to each other via
some dimension 4, 7, and 10 operators; the latter two will be suppressed by powers of a dimensionful scale
$\Lambda_3$. The interactions become strong at the energy scale $E\sim \Lambda_3$.
Nevertheless, we will show that the special structure  of interactions in
this theory guarantees that the operators presented in the tree-level Lagrangian
do not get renormalized at any order in perturbation theory.

The (non-canonically normalized) Lagrangian of the above-described theory reads as follows
\cite {deRham:2010ik}:
\beq
\mathcal{L}=-\frac{1}{2}
h^{\mu\nu}\mathcal{E}^{\alpha\beta}_{\mu\nu} h_{\alpha\beta}+
 h^{\mu\nu}\sum_{n=1}^3 \frac{a_{n}}{\Lambda^{3(n-1)}_3} X^{(n)}_{\mu\nu}\!\(\Pi\),
\label{lagr1}
\eeq
where $\mathcal{E}^{\alpha\beta}_{\mu\nu}$ is the Einstein operator, so that the first term denotes the quadratic Einstein-Hilbert contribution. The dimensionless coefficients $a_n$ are
tree-level free parameters (we will fix $a_1=-1/2$ below for a definite normalization of the scalar kinetic term) and the three $X$'s are explicitly given by the following expressions in terms of $\Pi\mn=\p_\mu\p_\nu \pi$ and the Levi-Civita symbol $\varepsilon^{\mu\nu\alpha\beta}$
\ba
X^{(1)}_{\mu\nu}\(\Pi\)&=&{\varepsilon_{\mu}}^{\alpha\rho\sigma}
{{\varepsilon_\nu}^{\beta}}_{\rho\sigma}\Pi_{\alpha\beta}, \quad  \nonumber \\
X^{(2)}_{\mu\nu}\(\Pi\)&=&{\varepsilon_{\mu}}^{\alpha\rho\gamma}
{{\varepsilon_\nu}^{\beta\sigma}}_{\gamma}\Pi_{\alpha\beta}
\Pi_{\rho\sigma}, \nonumber \\
X^{(3)}_{\mu\nu}\(\Pi\)&=&{\varepsilon_{\mu}}^{\alpha\rho\gamma}
{{\varepsilon_\nu}^{\beta\sigma\delta}}\Pi_{\alpha\beta}
\Pi_{\rho\sigma}\Pi_{\gamma\delta}\, .
\label{Xs}
\ea
$X^{(1,2,3)}$ are respectively linear, quadratic and cubic in $\p^2 \pi$, so that
the action involves operators up to quartic order in the fields.

The symmetries of the theory include: (a) linearized diffeomorphisms, $h\mn\to h\mn+\p_{(\mu}\xi_{\nu)}$, which represent an exact symmetry of the full non-linear action (\ie including the interactions $h^{\mu\nu}X^{(n)}\mn$), and (b) (global) field-space Galilean transformations, $\pi\to\pi+b_\mu x^\mu+b$ . The first of these is a symmetry up to a total derivative.

Although the interactions involve two derivatives on the scalar field $\pi$, the theory, defined by \eqref{lagr1} is ghost-free \cite{deRham:2010ik,deRham:2010kj}: it propagates exactly 2 polarizations of the massless tensor field and exactly one massless scalar; it thus represents a nontrivial example of
a model with a non-topologically interacting spin-2  and spin-0 fields.

We will show in the next section that the operators of this theory remain
protected against quantum corrections to all orders in perturbation theory,
despite the existence of non-trivial
interactions governed by the scale $\Lambda_3$. Technically, this
non-renormalization is due
to the specific structure of the interaction  vertices:  they  contain
two derivatives per scalar line, all contracted
by the epsilon tensors. Then, it is not too difficult to show,
as done  in the next section, that the loop diagrams  cannot induce
any renormalization of the tree-level terms in \eqref{lagr1}.
Conceptually, the non-renormalization appears because the tree-level interactions in the Lagrangian are diff invariant up to total derivatives only;
on the other hand, the variations of the Lagrangian w.r.t.
fields in this theory are exactly diff invariant;
therefore, no Feynman diagram  can generate operators that
would not be diff invariant, and the original operators that are diff
invariant only up to total derivatives
stay unrenormalized\footnote{This is similar to non-renormalization of the Galileon
operators \cite {Luty:2003vm, Nicolis:2004qq,Nicolis:2008in}, with diff invariance replaced by
galilean invariance;  we thank Kurt Hinterbichler for useful discussions on these
points.}.

In a conventional approach that would regard \eqref{lagr1} as an effective
field theory below the scale $\Lambda_3$,  there would  be new
terms induced by quantum loops, in addition to
the non-renormalizable terms already present in \eqref{lagr1}.
Let us consider one-loop terms
in the  1PI action. These are produced by an infinite number of one-loop
diagrams with external $h$ and/or  $\pi$ lines.  The diagrams contain
power-divergent terms,  the log-divergent pieces,  and finite terms.
The power-divergent terms  are arbitrary, and cannot be fixed without
the knowledge of the UV completion.  For instance,  dimensional
regularization would set these terms to zero. Alternatively,
one could use any other regularization, but  perform subsequent
subtraction so that the net result in the 1PI action is zero.

In contrast, the log divergent terms are uniquely determined:
they give rise to nonzero imaginary parts of
various amplitudes, such as the one depicted on Fig.1; the latter
determine the forward scattering cross sections
through the optical theorem. Therefore, these pieces  would have
to be included in the 1PI action.

All the induced terms in the 1PI action would appear suppressed
by the scale $\Lambda_3$, since the latter is  the only scale in the
effective field theory approach (including the scale of the UV cutoff).
Moreover, due to the same specific structure of the interaction
vertices that guarantees non-renormalization of \eqref{lagr1},
the  induced terms will have to have more derivatives per field than the
unrenormalized terms. Therefore at low energies, formally defined
by  the condition $(\partial/\Lambda_3)\ll 1$, the tree-level terms will
dominate over the induced terms with the same  number of fields, as well
as over the induced terms with a greater number of fields and derivatives.
This property clearly separates the unrenormalized terms from the
induced ones, and shows that the theory \eqref{lagr1} is a good effective
field theory below the scale $\Lambda_3$.

We now move to the discussion of  how classical sources enter the above picture.
As we will show  below, there are similarities to DGP \cite{Dvali:2000hr} and
Galileon \cite{Nicolis:2008in} theories, but there is also an
additional important ingredient that is specific to  the present theory
\eqref{lagr1}. We will present the discussion for the simplest case $a_3=0$,
when the Lagrangian can be explicitly diagonalized \cite {deRham:2010ik},
and will show in the next section that setting $a_3=0$ is technically natural.
Some of the novel qualitative features readily apply even when $a_3 \neq 0$,
however, in this case there are differences too, as we will briefly discuss below.

To this end, it is helpful to perform the field redefinition $h\mn \to h\mn + \pi \eta\mn$
and rewrite the two nonlinear interactions of the $a_3 =0$ theory as follows:
\beq
{a_2 \over \Lambda_3^3} \left ( 2G^{\mu\nu}(h) \partial_\mu \pi \partial_\nu \pi  +
3\square \pi (\partial_\mu \pi)^2 \right ).
\label{two}
\eeq
The second term in \eqref {two} is what appears in DGP (and hence in the
cubic Galileon theory). Since in this basis
the  $\pi$ field  couples to the  trace of the stress-tensor as $\pi T/M_{\rm Pl}$,
the nonlinear  term  $\square \pi (\partial\pi)^2$
gives  rise to the conventional Vainshtein mechanism: for a source of mass
$M_s\gg M_{\rm Pl}$, below the Vainshtein radius, $r_* =\({M_s/ M_{\rm Pl}}\)^{1/3}{\Lambda^{-1}_3}$,
the classical value of the $\pi$ field is severely suppressed (as compared
to its value in the linear theory) due to the fact that
in this regime $\partial^2 \pi /\Lambda_3^3 \gg1$ \footnote {The same applies to
time-dependent sources, for which, an additional scale  due to the time dependence
enters the Vainshtein radius \cite{deRham:2012fw}.}.
The novelty here is the first term in \eqref{two}: this term  dominates over
the second one both inside  and outside the source (but still inside the Vainshtein radius).
Outside the source it amounts of having a quartic Galileon in the theory,
and since its phenomenology is well-known \cite {Nicolis:2008in},
we will not discuss it in detail here. However, inside the source
we get $G_{\mu\nu}(h) \simeq T_{\mu\nu}/M_{\rm Pl}$, with a good accuracy.
Hence, in this region the $\pi$ field gets an additional kinetic term,
and the full quadratic term for it can be written as follows:
\beq
- \partial^\mu \pi \partial^\nu\pi \left (\eta_{\mu\nu} -
2 a_2 {T_{\mu\nu}\over \Lambda_3^3 M_{\rm Pl}}\right).
\label{kin}
\eeq
Thus, for a negative value of $a_2$ one gets a
classical renormalization of the  $\pi$ kinetic term. To appreciate how big
this renormalization is we note that the scale  in the denominator of the
second  term in \eqref {kin} is $M_{\rm Pl}^2 m^2$; for a graviton mass
comparable with the Hubble parameter $H_0$, this is of the order of the critical
density of the present day Universe. For instance, taking the Earth atmosphere as a
source, we get for the kinetic and gradient terms:
\beq
(1+ |a_2| 10^{26})(\partial_0\pi)^2 - (1+|a_2| 10^{14})(\partial_j\pi)^2\,.
\label{Zs}
\eeq
For higher density/pressure sources, such as the Earth itself,
or for any Earthly measuring device, we get even higher
factors of the order $10^{30}$ and $10^{18}$,
respectively for the kinetic and gradient terms.

Thanks to these new couplings, the strength
of the  interactions of the $\pi$ fluctuations above the classical
source, $\delta \pi = \pi - \pi_{\rm cl}$,  changes qualitatively.
Recall that in the DGP and the standard Galileon theories, the regime of  validity of the classical
solutions can be meaningfully established in the full quantum effective
theory due to the strong classical renormalization of the scalar kinetic terms
via the Vainshtein mechanism \cite{Nicolis:2004qq}
\footnote{The right procedure is to first solve
for a classical scalar profile in the presence of a source, and then calculate quantum
corrections. Of course the opposite
order should also give the same result once done correctly,
however in the latter case one would have to resum  quantum
corrections enhanced by large classical terms.}. Here we get an additional strong classical
renormalization of the kinetic term for the fluctuations $\delta \pi$
\beq
- \partial^\mu \delta \pi \partial^\nu \delta \pi \left ( Z^V_{\mu\nu} +
|a_2| Z^T_{\mu\nu} \right),
\label{kin2}
\eeq
where the first term in \eqref {kin2} is due to the Vainshtein mechanism,
which gives rise to a large $M_s$-dependent factor $Z^V\sim a_2 (r_*/R_{\rm Earth})$,
while the second one is  due to the above-mentioned novel coupling (the first term in \eqref {two}).

Furthermore, following  Ref. \cite  {Nicolis:2004qq}, the 1PI action can be organized
(using some reasonable assumptions about the UV theory) so that the local strong coupling
scale determining the interactions of the fluctuations
$\delta \pi$ schematically reads as follows:  $$\Lambda_{eff} (x) \equiv (Z^V +
|a_2| Z^T)^{1/3} \Lambda_3.$$
Very often $Z^T\gg Z^V$, and therefore $Z^T$ -- although localized in the source --
should be taken into account when and if bounds are imposed on the graviton mass from the
existence of this strong scale. For instance, as argued in \cite {Nicolis:2008in}
for the quartic Galileon, the angular part of the quadratic term for the fluctuations
is not enhanced by $Z^V$, presenting a challenge; luckily,
the enhancement due to $Z^T$ removes  this issue in the theory at hand.
Moreover, the $Z^T$-enhancement is present irrespective whether $a_3$ is chosen
to be zero or not -- it is solely defined by a nonzero $a_2$.
Regretfully, this effect has not been taken into account in Ref.~\cite{Burrage:2012ja}, and
the bounds on the graviton mass obtained there will have to be reconsidered ~\cite{Toappear}.

In addition to what we discussed above, there are additional  subtleties
when $a_3\neq 0$. In this case the Lagrangian \eqref {lagr1}
cannot be diagonalized by any local field redefinition \cite {deRham:2010ik}.
Hence, the nonlinear mixing term $h^{\mu\nu}X^{(3)}_{\mu\nu}$ will be present, no matter what.
Insertions of this vertex into quantum loops will generate higher powers and/or
derivatives  of the Riemann tensor $R_{\mu\alpha\nu\beta}$,
as well as mixed terms  between the Riemann tensor (with or without derivatives)
and derivatives of $\pi$. In a theory without sources all these terms
will be suppressed by $\Lambda_3$, again representing a good effective field
theory below this scale.

However, with classical sources included, there should appear a $Z$-factor supression
of the  terms containing  $R_{\mu\alpha\nu\beta}$, due to the fact that  on
nontrivial backgrounds of classical sources there will be a large quadratic
mixing  between  fluctuations of $h$ and $\pi$, and the latter has a
large kinetic term due to the $Z$ factor as discussed above.
All this will be discussed in Ref. \cite {Toappear}.

Last but not least, we note that the Lagrangian \eqref{lagr1} is not a garden-variety non-renormalizable model,
as it is clear from the above discussions, and there may be a diagram resummation approach
to the strong coupling issue, or perhaps a dual formulation along the lines of
Ref.~\cite{Gabadadze:2012sm}.  However, in the present work we adopted a conventional
low-energy effective field theory approach.

\subsection{Relation to massive gravity}

Interestingly, the action given in \eqref{lagr1} appears in a certain limit of a recently proposed class of theories of massive gravity, free of the Boulware-Deser (BD)
ghost \cite{Boulware:1973my}.
% CdR: We mention the application of massive gravity earlier in the introduction, and i think this sentence cuts the flow of the argument here, reason why i have commented it.
%
%Massive gravity is perhaps one of the most promising (and by far the most conservative) examples of large scale modified theories of gravity, with a potential of describing the accelerated expansion of the late-time Universe without the need to fine-tune the cosmological constant.
A two-parameter family of such theories has been proposed in \cite{deRham:2010ik,deRham:2010kj}. The theory has been shown to be free of the BD ghost perturbatively in \cite{deRham:2010kj}, at the full non-linear level in the Hamiltonian formalism in \cite{Hassan:2011hr,Hassan:2011ea}, and covariantly around any background in \cite{Mirbabayi:2011aa} (see also \cite{deRham:2011rn,deRham:2011qq,Hassan:2012qv} for a complementary analysis in the \stu and helicity languages, and \cite{Hinterbichler:2012cn} for a proof in the first order formalism).

This class of theories provides a promising framework for tackling the cosmological constant problem, given that the graviton mass $m$ can be tuned to be around the Hubble scale today. Such a tuning of $m$ with respect to the theoretically expected vacuum energy is of the same order as that of the conventional cosmological constant, $m^2 \lesssim 10^{-120} \mpl^2$; however, unlike the tuning of the cosmological constant, it is anticipated to be technically natural. The reason for this lies in the fact that in the $m\to 0$ limit we recover General Relativity, which, being a gauge theory, is fully protected from a quantum-mechanically induced graviton mass.

There is however a possible loophole in this reasoning: the $m\to0$ limit is obviously discontinuous in the number of gravitational degrees of freedom and the presence of these extra polarizations for $m\ne 0$ deserves a special treatment in the context of naturalness \footnote{Nevertheless, this  does not mean that the physical predictions of the theory are discontinuous. As mentioned above, the presence of the Vainshtein mechanism in this model \cite{Koyama:2011xz, Koyama:2011yg, Chkareuli:2011te, Sbisa:2012zk}, as well as general \cite{Babichev:2010jd} extensions of the Fierz-Pauli theory make most of the physical predictions identical to that of GR in the massless limit. }.

In particular, the theory \eqref{lagr1} emerges as the leading part of the ghost-free massive gravity action
%\footnote{A similar decoupling limit has been found in \cite{deRham:2011ca} for a recently pr%oposed three-dimensional New Massive Gravity \cite{Bergshoeff:2009hq}.},
describing the interactions of the helicity-2 and helicity-0 polarizations of the graviton in the limit
\beq
m\to 0,\qquad  \mpl\to \infty, \qquad \Lambda_3\equiv (\mpl m^2)^{1/3}=\text{finite}.
\label{declim}
\eeq
\begin{comment}
As an alternative approach, one can rely on the decoupling limit to infer that the additional degrees of freedom do decouple in the massless limit, and the local symmetry is indeed recover as $m\to 0$.
\end{comment}
Beyond this limit, the free parameters of the theory are expected to be renormalized, albeit by an amount that should vanish in the limit \eqref{declim}. As a result, quantum corrections to the three defining parameters of the full theory (namely the mass $m$ and the two free coefficients $a_{2,3}$) are strongly suppressed. In particular, the graviton mass receives a correction proportional to itself (with a coefficient that goes as $\delta m^2/m^2\sim \(m/\mpl \)^{2/3}$), thus establishing the technical naturalness of the theory.

One should stress at this point that technical naturalness is not an exclusive property of ghost-free massive gravity. Even theories with the BD ghost, can be technically natural, satisfying the $\delta m^2 \propto m^2$ property \cite{ArkaniHamed:2002sp}. Besides the fact that the latter theories are unacceptable,
there are  two important distinctions between the
theories with and without BD ghosts. These crucial distinctions can be formulated in the decoupling
limit, which occurs at a much lower energy scale, $\Lambda_5 = (m^4M_{\rm Pl})^{1/4} \ll \Lambda_3$,
if the theory propagates a BD ghost. In the latter case the classical part of the decoupling limit is
not protected by a non-renormalization theorem. As a consequence:
(a) quantum corrections in ghost-free theories are significantly suppressed with respect to those
in the theories with the BD ghost, and (b) unlike a generic massive gravity, the non-renormalization guarantees that \textit{any relative tuning of the parameters in the ghost-free theories, that is
$m,a_2,a_3$, is technically natural}. The latter property makes any relation between the free coefficients of the theory stable under quantum corrections \footnote{For example, a particular ghost-free theory with the decoupling limit, characterized by the vanishing of all interactions in \eqref{lagr1} has been studied due to its simplicity (\eg see Ref.~\cite{Buchbinder:2012wb} for one-loop divergences in that model). The non-renormalization of ghost-free massive gravity in this case guarantees that such a vanishing of the classical scalar-tensor interactions holds in the full quantum theory as well.}.

The rest of the paper is organized as follows. We show in Sec.~\ref{sec:NRtheorem} that the interactions of the scalar-tensor theory, defined in \eqref{lagr1} do not receive quantum corrections to any order in perturbation theory. Identifying the latter theory with the decoupling limit of massive gravity, we discuss the implications of such Renormalization Group (RG) invariance of relevant parameters in the given limit for the full theory, in Sec.~\ref{sec:MG} showing explicitly that quantum corrections to the graviton mass and the two free parameters of the potential are significantly suppressed. Finally, we conclude in Sec.~\ref{sec:Conclusion}.

\section{The Non-Renormalization Theorem}
\label{sec:NRtheorem}

In this section we present the non-renormalization argument for a class of scalar-tensor theories, defined by the Lagrangian \eqref{lagr1}. In particular, we will
show that the two parameters $a_{2,3}$ do not get renormalized, and that there
is no wave function renormalization for the spin-2 field $h\mn$.

Using the antisymmetric structure of these interactions, we can follow roughly the same arguments as for Galileon theories to show the RG invariance of these parameters, \cite{Luty:2003vm}. The only possible difference may emerge due to the gauge invariance $h\mn\to h\mn+\p_{(\mu}\xi_{\nu)}$, and consequently the necessity of gauge fixing for the tensor field. Working in {\it e.g.} the de Donder gauge, the relevant modification of the arguments is trivial: gauge invariance is Abelian, so the corresponding Faddeev-Popov ghosts are free and do not affect the argument in any way. Moreover, the gauge fixing term changes the graviton propagator, but as we shall see below, all the arguments that follow solely depend on the special structure of vertices and are hence independent of the exact structure of the propagator. With these arguments in mind, one can thus proceed with the proof of the non-renormalization of the theory without being affected by gauge invariance.

The scalar $\pi$ only appears within interactions/mixings with the spin-2 field in \eqref{lagr1}. In order to associate a propagator with it, we have to diagonalize the quadratic lagrangian by eliminating the $h^{\mu\nu}X\mn^{(1)}(\Pi)$ term.  Such a diagonalization gives rise to a kinetic term for $\pi$, as well as additional scalar self-interactions of the Galileon form \cite{deRham:2010ik},
\ba
\mathcal{L}=-\frac{1}{2}
h^{\mu\nu}\mathcal{E}^{\alpha\beta}_{\mu\nu} h_{\alpha\beta}+\frac{3}{2}\pi\Box\pi+
 \(h^{\mu\nu}+\pi \eta^{\mu\nu}\)\sum_{n=2}^3 \frac{a_{n}}{\Lambda^{3(n-1)}_3} X^{(n)}_{\mu\nu}\(\Pi\)\,,
\label{lagr01}
\ea
(here the interactions of the form $\pi X^{(n)}(\Pi)$ are nothing else but the cubic and quartic Galileons).
% in one of their incarnations).
%while diagonalizing the next mixed term $h^{\mu\nu}X\mn^{(2)}[\Pi]$ will give the rest of the Galileon interactions.
%

\begin{figure}[t]
\begin{center}
\includegraphics[height=2.8in,width=4in,angle=0]{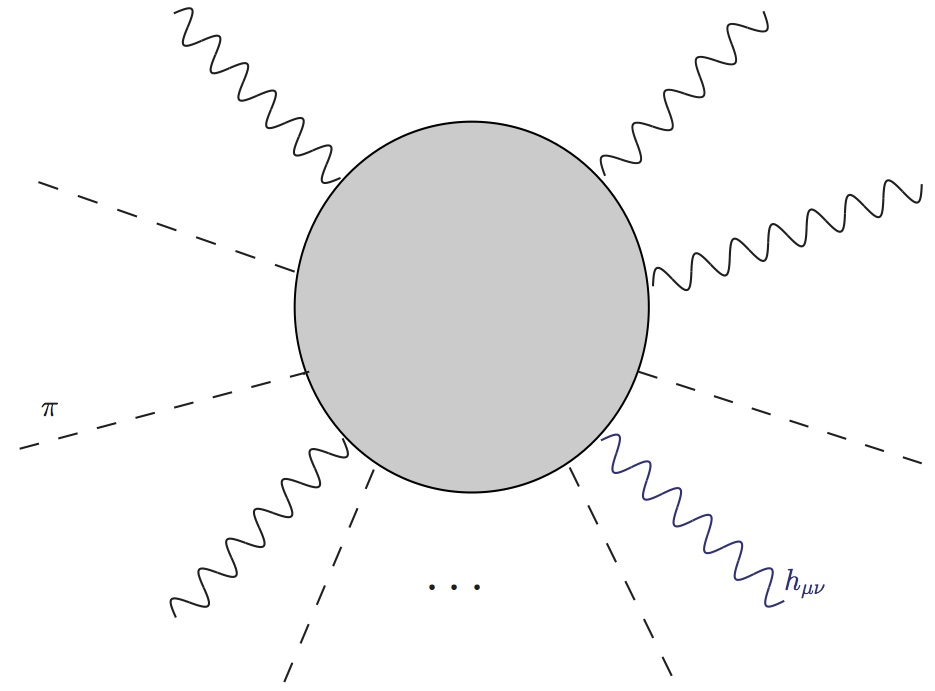}
\caption{An arbitrary 1PI diagram with gravitational degrees of freedom in the loop.}
\end{center}
\end{figure}

In the special case when the parameter $a_3$ vanishes, all scalar-tensor interactions are redundant and equivalent to pure scalar Galileon self-interactions. This can be seen through the field redefinition (under which the S-matrix is invariant) $h\mn=\tilde h\mn +\pi \eta\mn -\frac{2a_2}{\Lambda^3_3} \p_\mu \pi\p_\nu \pi$. We then recover a decoupled spin-2 field, supplemented by the Galileon theory for the scalar of the form
\ba
\L_{\rm Gal}= -\frac{1}{2}\sum_{n=0}^{2}\frac{b_n}{\Lambda^{3n}_3}X^{(n)}\mn\(\Pi\)\p^\mu \pi \p^\nu \pi\,,
\ea
where the Galileon coefficients $b_n$ are in one-to-one correspondence with $a_{n}$ and $X^{(0)}\mn \equiv \eta\mn$. The non-renormalization of the theory \eqref{lagr1} then directly follows from the analogous property of the Galileons. For $a_3\neq 0$, such a redefinition is however impossible \cite{deRham:2010ik}\footnote{This can be understood by noting that the $h^{\mu\nu}X^{(3)}\mn$ coupling encodes information about the linearized Riemann tensor for $h\mn$, which can not be expressed through $\pi$ on the basis of the lower-order equations of motion \cite{Chkareuli:2011te}.}.

We will now show that, similarly to what happens in the pure Galileon theories,  any external particle comes along with at least two derivatives acting on it in the 1PI action, hence establishing the non-renormalization of the operators present in \eqref{lagr1}. Of course, we keep in mind that these operators are merely the leading piece of the full 1PI action, which features an infinite number of additional higher derivative terms. They however are responsible for most of the phenomenology that the theories at hand lead to, making the non-renormalization property essential.

Consider an arbitrary 1PI diagram, such as the one depicted in Fig.~1. All vertices in \eqref{lagr01} have one field without a derivative, while all the rest come with two derivatives acting on them. Any external leg, contracted with a field with two derivatives in a vertex, obviously contributes to an operator with two derivatives on the field in the 1PI effective interaction, so if all the external legs were of that kind, this would lead to an operator of the form $\p^{2j}(\p^2 \pi)^k(\p^2 h\mn)^\ell$, with $j,k, \ell>0$. The only possibility of generating an operator with fewer derivatives on some of the fields comes from contracting fields without derivatives in vertices with external states.
For example, in the interaction $V=h^{\mu\nu}X\mn^{(2)}(\Pi) \sim h^{\mu\nu} \varepsilon_{\mu}^{\;\;\alpha\rho\gamma} \varepsilon_{\nu\;\;\;\;\gamma}^{\;\;\beta\sigma}\Pi_{\alpha\beta}\Pi_{\rho\sigma}$, the spin-2 field comes without derivatives, so let us look at an external $h\mn$ leg coming out of this vertex in an arbitrary 1PI graph, while letting the other two $\pi$-particles from this vertex run in the loop (of course, all of this reasoning will equivalently apply to any other vertex, such as $h^{\mu\nu}X\mn^{(3)}(\Pi)$, or $\pi X^{(2,3)}(\Pi) $). Let us denote the external, spin-2 momentum by $p_\mu$, while the momenta corresponding to the two $\pi$-particles in the loop are $k_\mu$ and $(p+k)_\mu$ respectively. The contribution of this vertex to the graph is given as follows
\ba
i\mathcal M
 \propto i \int \frac{\d^4 k}{(2\pi)^4}\mathcal{G}_k\mathcal{G}_{k+p}\  \epsilon^{*\mu\nu} \varepsilon_{\mu}^{\;\;\alpha\rho\gamma}
\varepsilon_{\nu\;\;\;\;\gamma}^{\;\;\beta\sigma}k_\alpha k_\beta (p+k)_\rho (p+k)_\sigma \cdots
\ea
where the Feynman propagator is denoted by $\mathcal{G}_k\equiv\frac{i}{k^2-i\epsilon}$ and $\epsilon^{*\mu\nu}$ is the spin-2 polarization tensor, while the ellipses encode information about the rest of the diagram. Now, the key observation is that the term independent of the external momentum $p$, as well as the term linear in it both cancel due to antisymmetric structure of the vertex. Hence, the only non-vanishing term involves two powers of the external spin-2 momentum $p_\rho p_\sigma$
\ba
i\mathcal M
 \propto i\epsilon^{*\mu\nu} \varepsilon_{\mu}^{\;\;\alpha\rho\gamma}
\varepsilon_{\nu\;\;\;\;\gamma}^{\;\;\beta\sigma}p_\rho p_\sigma \int\frac{\d^4k}{(2\pi)^4}\mathcal{G}_k\mathcal{G}_{k+p}k_\alpha k_\beta \cdots~,
\ea
yielding at least two derivatives on the external helicity - 2 mode in the position-space.
Thus any external leg coming out of the $h^{\mu\nu}X^{(2)}\mn$ vertex will necessarily have two or more derivatives on the corresponding field in the effective action. The same is trivially true for the $\pi X^{(2)}\mn$ vertex (as one can simply substitute $h\mn$ by $\pi$ in the above discussion).

Similarly, if the external leg is contracted with the derivative-free field in vertices $h^{\mu\nu}X^{(3)}\mn$ and $\pi X^{(3)}\mn$, their contribution will always involve the external momentum $p_\mu$ and the loop momenta $k_\mu$ and $k'_\mu$ with the following structure,
\ba
i\mathcal M
 &\propto & i \int \frac{\d^4 k \d^4 k'}{(2\pi)^8}\mathcal{G}_k\mathcal{G}_{k'}\mathcal{G}_{k+k'+p}
 f^{\mu\nu} \varepsilon_{\mu}^{\;\;\alpha\rho\gamma}
\varepsilon_{\nu}^{\;\;\beta\sigma\delta}k_\alpha k_\beta k'_\rho k'_\sigma (p+k+k')_\gamma (p+k+k')_\delta \cdots \nn \\
 &\propto & i  f^{\mu\nu} \varepsilon_{\mu}^{\;\;\alpha\rho\gamma}
\varepsilon_{\nu}^{\;\;\beta\sigma\delta} p_\gamma p_\delta  \int \frac{\d^4 k \d^4 k'}{(2\pi)^8}\mathcal{G}_k\mathcal{G}_{k'}\mathcal{G}_{k+k'+p}\   k_\alpha k_\beta k'_\rho k'_\sigma \cdots\,,\hspace{10pt}
\ea
where the contraction on the Levi-Civita symbols is performed with either the graviton polarization tensor $f^{\mu\nu}=\epsilon^{*\mu\nu}$ or with $f^{\mu\nu}=\eta^{\mu\nu}$ depending on whether we are dealing with the vertex $h^{\mu\nu}X^{(3)}\mn$ or $\pi X^{(3)}{}^\mu_\mu$. Similar arguments, as can be straightforwardly checked, lead to the same conclusion regarding the minimal number of derivatives on external fields for cases in which there are two external states coming out of these vertices (with the other two consequently running in the loops).

This completes the proof of the absence of quantum corrections to \begin{comment}
the scale $\Lambda_3$ and
\end{comment}
the two parameters $a_{2,3}$, as well as to the spin-2 kinetic term and the scalar-tensor kinetic mixing in the theory defined by \eqref{lagr1}.

\section{Massive Gravity and its Decoupling Limit}
\label{sec:MG}
For massive gravity  the action is a functional of the
metric $g_{\mu\nu}(x)$  and four spurious scalar fields
$\phi^a(x),~a=0,1,2,3$;  the latter are introduced to give a manifestly
diffeomorphism-invariant description \cite{Siegel:1993sk,ArkaniHamed:2002sp}.
One  defines a covariant tensor $H_{\mu\nu}$ as follows:
\beq
g_{\mu\nu} = \partial_\mu \phi^a
\partial_\nu \phi^b \eta_{ab} + H_{\mu\nu}\,,
\label{gH}
\eeq
where $\eta_{ab}={\rm diag}(-1,1,1,1)$ is the {\it field} space metric.
%
% CdR: i think the following sentence might be confusing, the reader might understand that massive gravity only has $H\mn$ as a dynamical field, which wouldnt be correct. I reformulated what i thought you had in mind.
%
% The first term on the r.h.s. is nothing but the Minkowski metric in the coordinate system defined by $\phi^a$'s.
%Hence, gravity in this formulation is described by
%the tensor  $H\mn$ propagating on Minkowski space.
In this formulation, the tensor $H\mn$ propagates on Minkowski space.
In the unitary gauge all the four  scalars $\phi^a(x)$ are frozen and equal to the corresponding space-time
coordinates,  $\phi^a(x)= x^\mu \delta_\mu^a$ and the tensor $H\mn$ coincides with the metric perturbation, $H\mn=h\mn$. However,
often it is helpful to use a non-unitary gauge in which the $\phi^a(x)$'s
are allowed to fluctuate.

A covariant Lagrangian density for massive gravity
can be written as follows,
\beq
\mathcal{L}=\frac{\mpl^2}{2} \sqrt{-g} \(R-\frac{m^2}{4}\,
\mathcal{U}(g,H)\)\,,
\label{L2}
\eeq
where $\mathcal{U}$ includes the mass,  and non-derivative
interaction  terms for $H_{\mu\nu}$ and  $g_{\mu\nu}$, while $R$ denotes the scalar curvature associated with the metric $g\mn$.

A necessary condition for the theory to be ghost free in the decoupling limit (DL)
is that  the potential  $\sqrt{-g}\ \mathcal{U}(g,H)$ be
a total derivative upon the field substitution $h_{\mu\nu}\equiv g_{\mu\nu} -\eta_{\mu\nu}=0,~~
\phi^a = \delta_\mu^a x^\mu - \eta^{a\mu}\partial_\mu\pi$  \cite{deRham:2010ik}.
With this substitution, the potential becomes a function of
$\Pi_{\mu\nu} \equiv \partial_\mu  \partial_\nu \pi $ and its various contractions with respect to the flat metric $\eta\mn$. The relevant terms can be constructed
straightforwardly by using the procedure outlined in Ref.~\cite{deRham:2010kj}.

In  any dimension there are only a finite number of
total derivative combinations, made out of $\Pi$, \cite{Nicolis:2008in}.  They are all
captured  by the recurrence relation \cite{deRham:2010ik}:
\beq
\label{Ldern}
\mathcal{L}_{\rm der}^{(n)}=-\sum_{m=1}^{n}(-1)^m\frac{(n-1)!}{(n-m)!}\,
[\Pi^{m}]\,\mathcal{L}^{(n-m)}_{\rm der}\,,
\eeq
with $\mathcal{L}^{(0)}_{\rm der}=1$ and $\mathcal{L}^{(1)}_{\rm der}=[\Pi]$.
This also guarantees that the sequence terminates,
\ie $\mathcal{L}^{(n)}_{\rm der}\equiv 0$,  for any $n\ge 5$ in four dimensions.
The list of all  nonzero total derivative terms starting with
the quadratic one reads as,
\beq
\label{L2der0}
\mathcal{L}^{(2)}_{\rm der}(\Pi) &=&[\Pi]^2-[\Pi^2]\,,\\
\label{L3der}
\mathcal{L}^{(3)}_{\rm der}(\Pi)&=&[\Pi]^3-3 [\Pi][\Pi^2]+2[\Pi^3]\,,\\
\label{L4der}
\mathcal{L}^{(4)}_{\rm der}(\Pi)&=&[\Pi]^4-6[\Pi^2][\Pi]^2+8[\Pi^3]
[\Pi]+3[\Pi^2]^2-6[\Pi^4]\, ,
\eeq
where we use the notation:  $[\Pi] \equiv {\rm Tr} \Pi^\mu_\nu $,
$[\Pi]^2 \equiv ({\rm Tr} \Pi^\mu_\nu )^2$, while
$ [\Pi^2 ] \equiv {\rm Tr} \Pi^\mu_\nu \Pi^\nu_\alpha$, with an obvious generalization
to terms of higher order in nonlinearity.

Then, as argued in \cite {deRham:2010kj}, the Lagrangian for massive gravity
that is automatically  ghost free  to all
orders in the DL  is obtained  by replacing the matrix elements
$\Pi^\mu_\nu $ in the total derivative terms \eqref{L2der0}-\eqref{L4der}
by the matrix elements  $\K^\mu_{~\nu}$,  defined as follows:
\beq
\label{Kmn}
\K^\mu_{~\nu} (g,H)\,=\,
\delta^\mu_\nu -\sqrt{\partial^\mu \phi^a \partial_\nu \phi^b \eta_{ab}}=\sqrt{\delta^\mu_{\nu}-H^{\mu}_{~\nu}}\,.
\eeq
Here, and everywhere below, the indices on $\K$ should be
lowered and raised with $g_{\mu\nu}$ and its inverse respectively.

This procedure defines the mass term, along with the interaction
potential in the Lagrangian density of
massive gravity \cite{deRham:2010kj}:
\begin{align}
\mathcal{L}&=\frac{\mpl^2}{2} \sqrt{-g} \left [ R + {m^2} \left(
\mathcal{L}^{(2)}_{\rm der}(\K) + \alpha_3 \mathcal{L}^{(3)}_{\rm der}(\K) +\alpha_4 \mathcal{L}^{(4)}_{\rm der}(\K)\right )  \right ]\nn \\
&=\frac{\mpl^2}{2} \sqrt{-g} \left [ R -\frac{m^2}{4} \left(H^{\mu\nu} H\mn-H^2+\dots \right)  \right ]\,.
\label{UUU}
\end{align}
Since all terms in (\ref {Ldern})  with $n\geq 5$
vanish identically,  by construction
all terms $\mathcal{L}^{(n)}_{\rm der}$ with $n\geq 5$
in (\ref {UUU}) are also zero. Hence, the most general Lagrangian density
(\ref {UUU}) has three free parameters, $m,\alpha_3$ and $\alpha_4$.

As is straightforward to see, Minkowski space $g\mn=\eta\mn$ with $\phi^a =x^a$ is a vacuum solution,
and the spectrum of the theory  (\ref{UUU})  contains a graviton of mass $m$;
the graviton also has  additional nonlinear interactions specified
by the action at hand.

The high-energy dynamics of the system is best displayed in the decoupling limit (DL), defined by \eqref{declim}. Being a direct analog of the nonlinear sigma model description of the high-energy limit of massive spin-1 theories, the decoupling limit of massive gravity features the five polarization states of the graviton, represented by  separate helicity states $0, \pm1, \pm2$. The helicity-2 mode $h\mn$ enters linearly in the decoupling limit, while the helicity-0 mode $\pi$ is fully non-linear\footnote{Here we denote the canonically normalized fields, obtained by $h\mn\to\frac{h\mn}{\mpl}$ and $\pi\to\frac{\pi}{\mpl m^2}$ by the same symbols.} (we will for the moment ignore the vector polarization and will comment on it below). The resulting DL theory takes precisely the form \eqref{declim}. It fully captures the most important features of massive gravity, such as the absence of the Boulware-Deser ghost \cite{deRham:2010ik} (which has been shown to generalize beyond the DL \cite{Mirbabayi:2011aa}), the existence of self-accelerating and screening solutions \cite{deRham:2010tw, Koyama:2011xz, D'Amico:2011jj, Gumrukcuoglu:2011ew}, etc. Moreover, the DL carries all the interactions that become relevant within the massless limit and provides a simple illustration of how the helicity-0 mode decouples from the rest of the gravitational sector for $m\to 0$ as an explicit realization of the Vainshtein mechanism. The limit at hand thus represents a powerful tool to study generic physical properties of the theory.
\begin{comment}The mere fact that the helicity-0 mode decouples in that limit is sufficient to deduce that four-dimensional diffeomorphism invariance is recovered as and hence the graviton mass can bear no quantum corrections as $m\to 0$. \end{comment}
Above we have uncovered a further interesting property of this limit: the non-renormalization theorem protecting the leading operators; we will see that this allows to extract information on quantum corrections to the full theory from simple DL arguments \cite{ArkaniHamed:2002sp}.

\subsection{Helicity-1 modes}

The scalar-tensor action given in \eqref{lagr1} does not include the DL interactions involving the helicity-1 modes $A_{\mu}$ of the massive graviton, defined through
\beq
\label{helicities}
\phi^a=\delta^a_\mu x^\mu-\eta^{a\mu} \(\frac{A_\mu}{\mpl m}+\frac{\p_\mu\pi}{\mpl m^2}\)\,,
\eeq
in \eqref{gH}.
So far, their precise form has only been found perturbatively (see for instance \cite{deRham:2010gu,Mirbabayi:2011aa,Tasinato:2012ze}). Schematically, to all orders they are given as follows \cite{Mirbabayi:2011aa}
\ba
\label{helicity1}
\L_{A}= - \frac 14 F^2+ F F\sum_{n>0}\frac{d_n}{\Lambda_3^{3n}} \Pi^{n}\,,
\ea
where $F$ denotes the field strength for $A_\mu$ and $d_n$ are constant coefficients.
These vertices can contribute to effective operators involving the helicity-0 mode, $\pi$. However, from the explicit form of these interactions it is manifest that every external $\pi$, originating from such a vertex will have at least two derivatives on it, in complete analogy to the case considered above. Taking into account the vector-scalar interactions of the form \eqref{helicity1} therefore does not change the non-renormalization properties of the scalar-tensor part of the action given in \eqref{lagr01}.

\begin{comment}
%
% CdR: i think the phrasing of the following sentence is misleading, if we really want to say something we can simply say that the coupling with external dynamical fields vanishes in the limit $\mpl \to \infty$. But this section on ```helicity-1" modes is not the place for that. So we can either emphasize this point before or after
%
In massive gravity, the DL non-renormalization also holds in the presence of external quantum matter fields, since loops involving such fields are suppressed by powers of $\mpl$ in these theories (these couplings are conventionally taken to be those of GR) and therefore vanish in the DL.
\end{comment}

\subsection{Implications for the full theory}

In this subsection we will comment on the implications of the above emergent DL non-renormalization property for the full theory. Below we will continue to treat massive gravity as an effective field theory
with a cutoff $\Lambda_3\gg m$.

We have established previously that in the DL the leading scalar-tensor part of the action does not receive quantum corrections in massive gravity: all operators generated by quantum corrections in the effective action have at least two extra derivatives compared to the leading terms, making the coefficients $a_i$ invariant under the renormalization group flow.  This in particular implies the absence of wave-function renormalization for the helicity-2 and helicity-0 fields in the DL. Moreover, the coupling with external matter fields goes as $\frac{1}{\mpl}h\mn T^{\mu\nu}$ and thus vanishes as $\mpl \to \infty$. The non-renormalization theorem is thus unaffected by external quantum matter fields.

The DL analysis of the effective action, much like the analogous nonlinear sigma models of non-Abelian spin-1 theories \cite{Vainshtein:1971ip}, provides an important advantage over the full treatment (see \cite{ArkaniHamed:2002sp} for a discussion of these matters.) In addition to being significantly simpler, the DL explicitly displays the relevant degrees of freedom and their (most important) interactions. In fact, as we will see below, we will be able to draw important conclusions regarding the magnitude of quantum corrections to the full theory based on the DL power counting analysis alone.

Now, whatever the renormalization of the specific coefficients $\alpha_i$ (and more generally, of any relative coefficient between terms of the form $[H^{\ell_1}]\cdots[H^{\ell_n}]$ in the graviton potential) is in the full theory \eqref{UUU}, it has to vanish in the DL, since $\alpha_i$ are in one-to-one correspondence with the unrenormalized DL parameters $a_i$. Let us work in the unitary gauge, in which $H\mn=h\mn$, and for example look at quadratic terms in the graviton potential. We start with an action, the relevant part of which (in terms of the so-far dimensionless $h\mn$) is
\beq
\label{lag}
\mathcal{L}\supset -\frac{1}{4} \mpl^2 m^2 \((1+c_1)~h^2\mn-(1+c_2)~h^2+\dots\),
\eeq
where $c_1$ and $c_2$ are generated by quantum corrections after integrating out a small Euclidean shell of momenta and indices are assumed to be contracted with the flat metric\footnote{We could as well assume that the full non-linear metric contracts indices, since the two cases are indistinguishable at the quadratic level.}. There is of course no guarantee that the two constants $c_{1,2}$ are equal, so they could lead to a detuning of the Fierz-Pauli structure and consequently to a ghost below the cutoff, unless sufficiently suppressed.
Returning to the \stu formalism, in terms of the canonically normalized fields
\beq
\label{scaling}
h\mn\to \frac{h\mn}{\mpl}~, \qquad \pi\to\frac{\pi}{\mpl m^2}
\eeq
the tree-level part (\textit{i.e.} the one without $c_1$ and $c_2$) of the above Lagrangian would lead to the following scalar-tensor kinetic mixing in the DL \eqref{lagr1}
\beq
\mathcal{L}\supset - h^{\mu\nu}\(\p_\mu\p_\nu\pi-\eta\mn\Box\pi\)+\dots~ .
\eeq
Now, from the DL analysis, we know that this mixing does not get renormalized. What does this imply for the renormalization of the graviton mass and the parameters of the potential in the full theory?

One immediate consequence of such non-renormalization is that in the decoupling limit, $c_1$ and $c_2$ both vanish. To infer the scaling of these parameters with $\mpl$, let us look at the scalar-tensor interactions that arise beyond the DL. They are of the following schematic form\footnote{We are omitting here the part containing
the helicity-1 interactions, which can uniquely be restored due to diff invariance of the
helicity-2+helicity-1 system,  and the $U(1)$ invariance of the helicity-1+helicity-0 system.}
\ba
\L=\sum_{n \ge 1,\, \ell\ge 0} \frac{f_{n,\ell}}{\Lambda^{3(\ell-1)}_3}h (\p^2 \pi)^\ell\(\frac{h}{\mpl}\)^n \,,
\ea
{\it i.e.}, they are all suppressed by an {\it integer} power of $\mpl^{-1}$ compared to vertices arising in the DL.  Then, judging from the structure of these interactions, generically the non-renormalization theorem for the classical scalar-tensor action should no longer be expected to hold
\begin{comment}
when considering the previous vertices present
\end{comment}
beyond the DL.

This implies that $c_{1}$ and $c_2$ generated by quantum corrections  are of the form
\ba
c_{1, 2} \sim \(\frac{\Lambda_3}{\mpl}\)^k\,,
\ea
with $k$ some positive integer $k\ge 1$, if the loops are to be cut off at the $\Lambda_3$ scale\footnote{In this analysis, the graviton mass $m$ is completely  absorbed into $\Lambda_3$, and nothing special happens at the scale $m$ as far as the strong coupling is concerned.} (the fact that $k$ needs to be an integer relies on the fact that the theory remains analytic beyond the DL.) Taking the worst possible case
({\it i.e.},  $k=1$), one can directly read off the magnitude of the coefficients $c_{1,2}$,
\beq
c_{1,2}\lsim \(\frac{\Lambda_3}{\mpl}\) \sim \(\frac{m}{\mpl}\)^{2/3}\,.
\eeq
In terms of the quantum correction to the graviton mass itself, this implies
\ba
\delta m^2 \lsim m^2 \(\frac{m}{\mpl}\)^{2/3} \,,
\ea
providing an explicit realization of technical naturalness for massive gravity.

One can extend these arguments to an arbitrary interaction in the effective potential. Consider a generic term of the following schematic form in the unitary gauge involving $\ell$ factors of the (dimensionless) metric perturbation
\beq
\mathcal{L}\supset \mpl^2 m^2 \sqrt{-g} ~ (\bar c+c)h^\ell~.
\eeq
Here indices are contracted with the full metric, $\bar c$ denotes the ``classical" coefficient of the given term obtained from \eqref{UUU}, and $c$ is its quantum correction. Our task is to estimate the magnitude of $c$ based on the non-renormalization of the DL scalar-tensor Lagrangian. Introducing back the \stu fields through the replacement $h\mn\to H\mn$, and recalling the definition of different helicities \eqref{helicities}, the quantum correction to the given interaction can be schematically written in terms of the various canonically normalized helicities as follows
\beq
\(1+\frac{h}{\mpl}+\dots\)^{1+\ell} \(\frac{h}{\mpl}+\frac{\p A}{\mpl m}+\frac{\p^2\pi}{\Lambda_3^3} +\frac{\p A \p^2\pi}{\mpl m \Lambda_3^3}+\frac{(\p A)^2}{\mpl^2 m^2}+\frac{(\p^2 \pi)^2}{\Lambda_3^6} \)^\ell .\nn
\eeq
The first parentheses denotes a schematic product of $\sqrt{-g}$ and $\ell$ factors of the inverse metric, needed to contract the indices. In the classical ghost-free massive gravity, the pure scalar self-interactions are carefully tuned to collect into total derivatives, projecting out the BD ghost. From the DL arguments, we know that quantum corrections do produce such operators, {\it e.g.} of the form $(\p^2\pi)^\ell$, suppressed by the powers of $\Lambda_3$. This immediately bounds the magnitude of the coefficient $c$ to be the same as for the $\ell=2$ case
\beq
c\lsim \(\frac{\Lambda_3}{\mpl}\)\sim \(\frac{m}{\mpl}\)^{2/3}~.
\label{bound}
\eeq
Indeed, for $c$ given by \eqref{bound}, we get $\mpl^2 m^2 c\sim \Lambda^4_3$ and the upper bound on $c$ is the same as that coming from the mass term renormalization.

\section{Discussion and Conclusions}
\label{sec:Conclusion}
% We have presented a non-renormalization theorem in a class of non-renormalizable scalar-tensor theories, relevant for infrared modifications of gravity.
% CdR: replaced the previous sentence with
We have presented a non-renormalization theorem in a special class of scalar-tensor theories, relevant for infrared modifications of gravity.

Although these theories feature irrelevant, non-topological interactions of a spin-2 field with a scalar, the couplings corresponding to these interactions do not get renormalized to any order in perturbation theory. This provides an interesting example of non-renormalization in non-supersymmetric theories with dimensionful couplings.

The scalar-tensor theories of this kind arise in the DL of the recently proposed models of ghost-free massive gravity. The emergent DL non-renormalization property, as we have seen, allows one to estimate the magnitude of quantum corrections to various parameters defining the full theory beyond any limit. In particular, one can show that setting an arbitrarily small graviton mass is technically natural. The significance of the DL theory is hard to overestimate: it unambiguously determines all the physical dynamics of the theory at distances $\Lambda^{-1}_{eff}\lsim r\lsim m^{-1}$, essentially capturing all physics at astrophysical and cosmological scales. Technical naturalness, along with a yet stronger non-renormalization theorem provide perfect predictivity of the theory, enforcing quantum corrections to play essentially no role at these scales. Moreover, dictated by various physical considerations, one frequently chooses to set certain relations between the two free parameters of the theory, $\alpha_3$ and $\alpha_4$. Non-renormalization in this case means, that such relative tunings of parameters, along with any physical consequences that these tunings may have, are not subject to destabilization via quantum corrections.

In this work, we have not made any assumptions regarding the UV completion of the ghost-free massive gravity. The special structure of the graviton potential might lead to a resummation of the infinite number of loop diagrams allowing to stay in the weakly-coupled regime without the need of invoking new dynamics. The properties of the DL, including non-renormalization, might be pointing towards such a simplification of the S-matrix at the apparent strong coupling scale $\Lambda_3$, which might become transparent in a certain alternative field basis \cite{Gabadadze:2012sm} (for other proposals for UV behavior, see \cite{Dvali:2010jz,*Dvali:2010ns}.)

We have not addressed the latter questions here and have presented a standard effective field theory interpretation of massive gravity. Given the effective theory at the scale $\Lambda_3$, we have shown that the couplings of the leading (decoupling limit) action are RG-invariant as the theory flows towards the infrared. Since it is precisely this part of the action that is responsible for most of the relevant physics at the astrophysical/cosmological scales, one arrives at rather powerful predictivity properties of the theory: (a) \emph{all the defining parameters of the theory are technically natural}; (b) \emph{moreover, any choice of relations between them is also technically natural: if one sets a relation at the scale $\Lambda_3$, it remains unchanged at any other lower scale.} This leads to the possibility to study the predictions of the classical theory at the scales at hand without ever worrying about quantum corrections.

We expect similar non-renormalization properties to hold in the recently proposed theory of quasi-dilaton massive gravity \cite{D'Amico:2012zv}.

\vskip.5cm

\bigskip
{\bf Acknowledgments}:
We would like to thank K.~Hinterbichler, R.~Rosen and A.~J.~Tolley for useful discussions. GG is supported by NSF grant PHY-0758032. DP is supported by the U.S. Department of Energy under contract No. DOE-FG03-97ER40546 and would like to thank the Center for Cosmology and Particle Physics at New York University for hospitality during the final stages of completion of the paper.

\bibliographystyle{utphys}
\addcontentsline{toc}{section}{References}
\bibliography{qcorr6}

\providecommand{\href}[2]{#2}\begingroup\raggedright\begin{thebibliography}{10}

\bibitem{deRham:2010tw}
C.~de~Rham, G.~Gabadadze, L.~Heisenberg, and D.~Pirtskhalava, ``{Cosmic
  Acceleration and the Helicity-0 Graviton},''
  \href{http://dx.doi.org/10.1103/PhysRevD.83.103516}{{\em Phys.Rev.}
  {\bfseries D83} (2011) 103516},
\href{http://arxiv.org/abs/1010.1780}{{\ttfamily arXiv:1010.1780 [hep-th]}}.
%%CITATION = ARXIV:1010.1780;%%.

\bibitem{'tHooft:1979bh}
G.~'t~Hooft, ``{Naturalness, chiral symmetry, and spontaneous chiral symmetry
  breaking},''
{\em NATO Adv.Study Inst.Ser.B Phys.} {\bfseries 59} (1980) 135.
%%CITATION = NASBD,59,135;%%.

\bibitem{Dimopoulos:1979es}
S.~Dimopoulos and L.~Susskind, ``{Mass Without Scalars},''
\href{http://dx.doi.org/10.1016/0550-3213(79)90364-X}{{\em Nucl.Phys.}
  {\bfseries B155} (1979) 237--252}.
%%CITATION = NUPHA,B155,237;%%.

\bibitem{Vainshtein:1972sx}
A.~Vainshtein, ``{To the problem of nonvanishing gravitation mass},''
\href{http://dx.doi.org/10.1016/0370-2693(72)90147-5}{{\em Phys.Lett.}
  {\bfseries B39} (1972) 393--394}.
%%CITATION = PHLTA,B39,393;%%.

\bibitem{Deffayet:2001uk}
C.~Deffayet, G.~Dvali, G.~Gabadadze, and A.~I. Vainshtein, ``{Nonperturbative
  continuity in graviton mass versus perturbative discontinuity},''
  \href{http://dx.doi.org/10.1103/PhysRevD.65.044026}{{\em Phys.Rev.}
  {\bfseries D65} (2002) 044026},
\href{http://arxiv.org/abs/hep-th/0106001}{{\ttfamily arXiv:hep-th/0106001
  [hep-th]}}.
%%CITATION = HEP-TH/0106001;%%.

\bibitem{Adams:2006sv}
A.~Adams, N.~Arkani-Hamed, S.~Dubovsky, A.~Nicolis, and R.~Rattazzi,
  ``{Causality, analyticity and an IR obstruction to UV completion},''
  \href{http://dx.doi.org/10.1088/1126-6708/2006/10/014}{{\em JHEP} {\bfseries
  0610} (2006) 014},
\href{http://arxiv.org/abs/hep-th/0602178}{{\ttfamily arXiv:hep-th/0602178
  [hep-th]}}.
%%CITATION = HEP-TH/0602178;%%.

\bibitem{Burrage:2011cr}
C.~Burrage, C.~de~Rham, L.~Heisenberg, and A.~J. Tolley, ``{Chronology
  Protection in Galileon Models and Massive Gravity},''
  \href{http://dx.doi.org/10.1088/1475-7516/2012/07/004}{{\em JCAP} {\bfseries
  1207} (2012) 004},
\href{http://arxiv.org/abs/1111.5549}{{\ttfamily arXiv:1111.5549 [hep-th]}}.
%%CITATION = ARXIV:1111.5549;%%.

\bibitem{deRham:2010ik}
C.~de~Rham and G.~Gabadadze, ``{Generalization of the Fierz-Pauli Action},''
  \href{http://dx.doi.org/10.1103/PhysRevD.82.044020}{{\em Phys.Rev.}
  {\bfseries D82} (2010) 044020},
\href{http://arxiv.org/abs/1007.0443}{{\ttfamily arXiv:1007.0443 [hep-th]}}.
%%CITATION = ARXIV:1007.0443;%%.

\bibitem{deRham:2010kj}
C.~de~Rham, G.~Gabadadze, and A.~J. Tolley, ``{Resummation of Massive
  Gravity},'' \href{http://dx.doi.org/10.1103/PhysRevLett.106.231101}{{\em
  Phys.Rev.Lett.} {\bfseries 106} (2011) 231101},
\href{http://arxiv.org/abs/1011.1232}{{\ttfamily arXiv:1011.1232 [hep-th]}}.
%%CITATION = ARXIV:1011.1232;%%.

\bibitem{Luty:2003vm}
M.~A. Luty, M.~Porrati, and R.~Rattazzi, ``{Strong interactions and stability
  in the DGP model},'' {\em JHEP} {\bfseries 0309} (2003) 029,
\href{http://arxiv.org/abs/hep-th/0303116}{{\ttfamily arXiv:hep-th/0303116
  [hep-th]}}.
%%CITATION = HEP-TH/0303116;%%.

\bibitem{Nicolis:2004qq}
A.~Nicolis and R.~Rattazzi, ``{Classical and quantum consistency of the DGP
  model},'' \href{http://dx.doi.org/10.1088/1126-6708/2004/06/059}{{\em JHEP}
  {\bfseries 0406} (2004) 059},
\href{http://arxiv.org/abs/hep-th/0404159}{{\ttfamily arXiv:hep-th/0404159
  [hep-th]}}.
%%CITATION = HEP-TH/0404159;%%.

\bibitem{Nicolis:2008in}
A.~Nicolis, R.~Rattazzi, and E.~Trincherini, ``{The Galileon as a local
  modification of gravity},''
  \href{http://dx.doi.org/10.1103/PhysRevD.79.064036}{{\em Phys.Rev.}
  {\bfseries D79} (2009) 064036},
\href{http://arxiv.org/abs/0811.2197}{{\ttfamily arXiv:0811.2197 [hep-th]}}.
%%CITATION = ARXIV:0811.2197;%%.

\bibitem{Dvali:2000hr}
G.~Dvali, G.~Gabadadze, and M.~Porrati, ``{4-D gravity on a brane in 5-D
  Minkowski space},''
  \href{http://dx.doi.org/10.1016/S0370-2693(00)00669-9}{{\em Phys.Lett.}
  {\bfseries B485} (2000) 208--214},
\href{http://arxiv.org/abs/hep-th/0005016}{{\ttfamily arXiv:hep-th/0005016
  [hep-th]}}.
%%CITATION = HEP-TH/0005016;%%.

\bibitem{deRham:2012fw}
C.~de~Rham, A.~J. Tolley, and D.~H. Wesley, ``{Vainshtein Mechanism in Binary
  Pulsars},''
\href{http://arxiv.org/abs/1208.0580}{{\ttfamily arXiv:1208.0580 [gr-qc]}}.
%%CITATION = ARXIV:1208.0580;%%.

\bibitem{Burrage:2012ja}
C.~Burrage, N.~Kaloper, and A.~Padilla, ``{Strong Coupling and Bounds on the
  Graviton Mass in Massive Gravity},''
\href{http://arxiv.org/abs/1211.6001}{{\ttfamily arXiv:1211.6001 [hep-th]}}.
%%CITATION = ARXIV:1211.6001;%%.

\bibitem{Toappear}
\text{To appear}.

\bibitem{Gabadadze:2012sm}
G.~Gabadadze, K.~Hinterbichler, and D.~Pirtskhalava, ``{Classical Duals of
  Derivatively Self-Coupled Theories},''
  \href{http://dx.doi.org/10.1103/PhysRevD.85.125007}{{\em Phys.Rev.}
  {\bfseries D85} (2012) 125007},
\href{http://arxiv.org/abs/1202.6364}{{\ttfamily arXiv:1202.6364 [hep-th]}}.
%%CITATION = ARXIV:1202.6364;%%.

\bibitem{Boulware:1973my}
D.~Boulware and S.~Deser, ``{Can gravitation have a finite range?},''
\href{http://dx.doi.org/10.1103/PhysRevD.6.3368}{{\em Phys.Rev.} {\bfseries D6}
  (1972) 3368--3382}.
%%CITATION = PHRVA,D6,3368;%%.

\bibitem{Hassan:2011hr}
S.~Hassan and R.~A. Rosen, ``{Resolving the Ghost Problem in non-Linear Massive
  Gravity},'' \href{http://dx.doi.org/10.1103/PhysRevLett.108.041101}{{\em
  Phys.Rev.Lett.} {\bfseries 108} (2012) 041101},
\href{http://arxiv.org/abs/1106.3344}{{\ttfamily arXiv:1106.3344 [hep-th]}}.
%%CITATION = ARXIV:1106.3344;%%.

\bibitem{Hassan:2011ea}
S.~Hassan and R.~A. Rosen, ``{Confirmation of the Secondary Constraint and
  Absence of Ghost in Massive Gravity and Bimetric Gravity},''
  \href{http://dx.doi.org/10.1007/JHEP04(2012)123}{{\em JHEP} {\bfseries 1204}
  (2012) 123},
\href{http://arxiv.org/abs/1111.2070}{{\ttfamily arXiv:1111.2070 [hep-th]}}.
%%CITATION = ARXIV:1111.2070;%%.

\bibitem{Mirbabayi:2011aa}
M.~Mirbabayi, ``{A Proof Of Ghost Freedom In de Rham-Gabadadze-Tolley Massive
  Gravity},''
\href{http://arxiv.org/abs/1112.1435}{{\ttfamily arXiv:1112.1435 [hep-th]}}.
%%CITATION = ARXIV:1112.1435;%%.

\bibitem{deRham:2011rn}
C.~de~Rham, G.~Gabadadze, and A.~J. Tolley, ``{Ghost free Massive Gravity in
  the St\'uckelberg language},''
  \href{http://dx.doi.org/10.1016/j.physletb.2012.03.081}{{\em Phys.Lett.}
  {\bfseries B711} (2012) 190--195},
\href{http://arxiv.org/abs/1107.3820}{{\ttfamily arXiv:1107.3820 [hep-th]}}.
%%CITATION = ARXIV:1107.3820;%%.

\bibitem{deRham:2011qq}
C.~de~Rham, G.~Gabadadze, and A.~J. Tolley, ``{Helicity Decomposition of
  Ghost-free Massive Gravity},''
  \href{http://dx.doi.org/10.1007/JHEP11(2011)093}{{\em JHEP} {\bfseries 1111}
  (2011) 093},
\href{http://arxiv.org/abs/1108.4521}{{\ttfamily arXiv:1108.4521 [hep-th]}}.
%%CITATION = ARXIV:1108.4521;%%.

\bibitem{Hassan:2012qv}
S.~Hassan, A.~Schmidt-May, and M.~von Strauss, ``{Proof of Consistency of
  Nonlinear Massive Gravity in the St\'uckelberg Formulation},''
  \href{http://dx.doi.org/10.1016/j.physletb.2012.07.018}{{\em Phys.Lett.}
  {\bfseries B715} (2012) 335--339},
\href{http://arxiv.org/abs/1203.5283}{{\ttfamily arXiv:1203.5283 [hep-th]}}.
%%CITATION = ARXIV:1203.5283;%%.

\bibitem{Hinterbichler:2012cn}
K.~Hinterbichler and R.~A. Rosen, ``{Interacting Spin-2 Fields},''
  \href{http://dx.doi.org/10.1007/JHEP07(2012)047}{{\em JHEP} {\bfseries 1207}
  (2012) 047},
\href{http://arxiv.org/abs/1203.5783}{{\ttfamily arXiv:1203.5783 [hep-th]}}.
%%CITATION = ARXIV:1203.5783;%%.

\bibitem{Koyama:2011xz}
K.~Koyama, G.~Niz, and G.~Tasinato, ``{Analytic solutions in non-linear massive
  gravity},'' \href{http://dx.doi.org/10.1103/PhysRevLett.107.131101}{{\em
  Phys.Rev.Lett.} {\bfseries 107} (2011) 131101},
\href{http://arxiv.org/abs/1103.4708}{{\ttfamily arXiv:1103.4708 [hep-th]}}.
%%CITATION = ARXIV:1103.4708;%%.

\bibitem{Koyama:2011yg}
K.~Koyama, G.~Niz, and G.~Tasinato, ``{Strong interactions and exact solutions
  in non-linear massive gravity},''
  \href{http://dx.doi.org/10.1103/PhysRevD.84.064033}{{\em Phys.Rev.}
  {\bfseries D84} (2011) 064033},
\href{http://arxiv.org/abs/1104.2143}{{\ttfamily arXiv:1104.2143 [hep-th]}}.
%%CITATION = ARXIV:1104.2143;%%.

\bibitem{Chkareuli:2011te}
G.~Chkareuli and D.~Pirtskhalava, ``{Vainshtein Mechanism In $\Lambda_3$ -
  Theories},'' \href{http://dx.doi.org/10.1016/j.physletb.2012.05.030}{{\em
  Phys.Lett.} {\bfseries B713} (2012) 99--103},
\href{http://arxiv.org/abs/1105.1783}{{\ttfamily arXiv:1105.1783 [hep-th]}}.
%%CITATION = ARXIV:1105.1783;%%.

\bibitem{Sbisa:2012zk}
F.~Sbisa, G.~Niz, K.~Koyama, and G.~Tasinato, ``{Characterising Vainshtein
  Solutions in Massive Gravity},''
  \href{http://dx.doi.org/10.1103/PhysRevD.86.024033}{{\em Phys.Rev.}
  {\bfseries D86} (2012) 024033},
\href{http://arxiv.org/abs/1204.1193}{{\ttfamily arXiv:1204.1193 [hep-th]}}.
%%CITATION = ARXIV:1204.1193;%%.

\bibitem{Babichev:2010jd}
E.~Babichev, C.~Deffayet, and R.~Ziour, ``{The Recovery of General Relativity
  in massive gravity via the Vainshtein mechanism},''
  \href{http://dx.doi.org/10.1103/PhysRevD.82.104008}{{\em Phys.Rev.}
  {\bfseries D82} (2010) 104008},
\href{http://arxiv.org/abs/1007.4506}{{\ttfamily arXiv:1007.4506 [gr-qc]}}.
%%CITATION = ARXIV:1007.4506;%%.

\bibitem{ArkaniHamed:2002sp}
N.~Arkani-Hamed, H.~Georgi, and M.~D. Schwartz, ``{Effective field theory for
  massive gravitons and gravity in theory space},''
  \href{http://dx.doi.org/10.1016/S0003-4916(03)00068-X}{{\em Annals Phys.}
  {\bfseries 305} (2003) 96--118},
\href{http://arxiv.org/abs/hep-th/0210184}{{\ttfamily arXiv:hep-th/0210184
  [hep-th]}}.
%%CITATION = HEP-TH/0210184;%%.

\bibitem{Buchbinder:2012wb}
I.~Buchbinder, D.~Pereira, and I.~Shapiro, ``{One-loop divergences in massive
  gravity theory},''
  \href{http://dx.doi.org/10.1016/j.physletb.2012.04.045}{{\em Phys.Lett.}
  {\bfseries B712} (2012) 104--108},
\href{http://arxiv.org/abs/1201.3145}{{\ttfamily arXiv:1201.3145 [hep-th]}}.
%%CITATION = ARXIV:1201.3145;%%.

\bibitem{Siegel:1993sk}
W.~Siegel, ``{Hidden gravity in open string field theory},''
  \href{http://dx.doi.org/10.1103/PhysRevD.49.4144}{{\em Phys.Rev.} {\bfseries
  D49} (1994) 4144--4153},
\href{http://arxiv.org/abs/hep-th/9312117}{{\ttfamily arXiv:hep-th/9312117
  [hep-th]}}.
%%CITATION = HEP-TH/9312117;%%.

\bibitem{D'Amico:2011jj}
G.~D'Amico, C.~de~Rham, S.~Dubovsky, G.~Gabadadze, D.~Pirtskhalava, {\em
  et~al.}, ``{Massive Cosmologies},''
  \href{http://dx.doi.org/10.1103/PhysRevD.84.124046}{{\em Phys.Rev.}
  {\bfseries D84} (2011) 124046},
\href{http://arxiv.org/abs/1108.5231}{{\ttfamily arXiv:1108.5231 [hep-th]}}.
%%CITATION = ARXIV:1108.5231;%%.

\bibitem{Gumrukcuoglu:2011ew}
A.~E. Gumrukcuoglu, C.~Lin, and S.~Mukohyama, ``{Open FRW universes and
  self-acceleration from nonlinear massive gravity},''
  \href{http://dx.doi.org/10.1088/1475-7516/2011/11/030}{{\em JCAP} {\bfseries
  1111} (2011) 030},
\href{http://arxiv.org/abs/1109.3845}{{\ttfamily arXiv:1109.3845 [hep-th]}}.
%%CITATION = ARXIV:1109.3845;%%.

\bibitem{deRham:2010gu}
C.~de~Rham and G.~Gabadadze, ``{Selftuned Massive Spin-2},''
  \href{http://dx.doi.org/10.1016/j.physletb.2010.08.043}{{\em Phys.Lett.}
  {\bfseries B693} (2010) 334--338},
\href{http://arxiv.org/abs/1006.4367}{{\ttfamily arXiv:1006.4367 [hep-th]}}.
%%CITATION = ARXIV:1006.4367;%%.

\bibitem{Tasinato:2012ze}
G.~Tasinato, K.~Koyama, and G.~Niz, ``{Vector instabilities and
  self-acceleration in the decoupling limit of massive gravity},''
\href{http://arxiv.org/abs/1210.3627}{{\ttfamily arXiv:1210.3627 [hep-th]}}.
%%CITATION = ARXIV:1210.3627;%%.

\bibitem{Vainshtein:1971ip}
A.~Vainshtein and I.~Khriplovich, ``{On the zero-mass limit and
  renormalizability in the theory of massive yang-mills field},''
{\em Yad.Fiz.} {\bfseries 13} (1971) 198--211.
%%CITATION = YAFIA,13,198;%%.

\bibitem{Dvali:2010jz}
G.~Dvali, G.~F. Giudice, C.~Gomez, and A.~Kehagias, ``{UV-Completion by
  Classicalization},'' \href{http://dx.doi.org/10.1007/JHEP08(2011)108}{{\em
  JHEP} {\bfseries 1108} (2011) 108},
\href{http://arxiv.org/abs/1010.1415}{{\ttfamily arXiv:1010.1415 [hep-ph]}}.
%%CITATION = ARXIV:1010.1415;%%.

\bibitem{Dvali:2010ns}
G.~Dvali and D.~Pirtskhalava, ``{Dynamics of Unitarization by
  Classicalization},''
  \href{http://dx.doi.org/10.1016/j.physletb.2011.03.054}{{\em Phys.Lett.}
  {\bfseries B699} (2011) 78--86},
\href{http://arxiv.org/abs/1011.0114}{{\ttfamily arXiv:1011.0114 [hep-ph]}}.
%%CITATION = ARXIV:1011.0114;%%.

\bibitem{D'Amico:2012zv}
G.~D'Amico, G.~Gabadadze, L.~Hui, and D.~Pirtskhalava, ``{Quasi-Dilaton: Theory
  and Cosmology},''
\href{http://arxiv.org/abs/1206.4253}{{\ttfamily arXiv:1206.4253 [hep-th]}}.
%%CITATION = ARXIV:1206.4253;%%.

\end{thebibliography}\endgroup

\end{document}